\documentclass[11pt]{article}

\usepackage{amsmath}

\newtheorem{theorem}{Theorem}

\begin{document}

\title{New solvable problems in the dynamics of a rigid body about a fixed
point in a potential field.}
\author{Yehia Hamad M.\\
Department of Mathematics, Faculty of Science, Mansoura University,\\
Mansoura 35516, Egypt\\
Email: hyehia@mans.edu.eg\\
Tel: +201001780325. Fax +20502246781}
\maketitle

\begin{abstract}
We determine the general form of the potential of the problem of motion of a
rigid body about a fixed point, which allows the angular velocity to remain
permanently in a principal plane of inertia of the body. Explicit solution
of the problem of motion is reduced to inversion of a single integral. A
several-parameter generalization of the classical case due to Bobylev and
Steklov is found. Special cases solvable in elliptic and ultraelliptic
functions of time are discussed.

\end{abstract}

\section{Introduction}

Integrable systems constitute a rare exception in mechanics. This is most
clearly manifested in the field of rigid body dynamics, where all known
integrable cases constitute few small tables, see e.g. \cite{lei}, \cite%
{yjpall} and \cite{bor}. Second comes particular solutions of equations of
motion of a rigid body in various settings. Those are solutions valid only
for certain particular sets of the initial position and angular velocity. In
the classical problem of motion of a rigid body about a fixed point in a
uniform gravity field there are eleven solutions of this type known after
authors of the 19th and the 20th centuries. All of them are collected in
table 1 below (in chronological order):

\begin{equation*}
\begin{tabular}{llllll}
\hline
Case & \ \ \ \ \ 1 & \ \ \ \ \ \ 2 & \ \ \ \ \ \ 3 & \ \ \ \ \ \ 4 & \ \ \ \
\ \ 5 \\ \hline
Au. & Hess & Staude & 
\begin{tabular}{l}
Bobylev- \\ 
Steklov%
\end{tabular}
& Goriatchev & Steklov \\ \hline
Year & 1890 & 1894 & 1896 & 1899 & 1899 \\ \hline
Ref. & \cite{hess} & \cite{stau} & \cite{bob, st} & \cite{gor} & \cite{stek}
\\ \hline
Case & \ \ \ \ \ 6 & \ \ \ \ 7 & \ \ \ \ \ 8 & \ \ \ \ \ 9,11 & \ \ \ \ 10
\\ \hline
Au. & Chaplygin & Kowalewski & Grioli & Dokshevich & 
\begin{tabular}{l}
Konosevich- \\ 
Pozdnyakovich%
\end{tabular}
\\ \hline
Year & 1904 & 1908 & 1947 & 1965, 1970 & 1968 \\ \hline
Ref. & \cite{chap} & \cite{kow} & \cite{grio} & \cite{dok1, dok2} & \cite%
{kono-poz} \\ \hline
\end{tabular}%
\end{equation*}

\begin{center}
\bigskip Table 1: Known particular solvable cases of the classical problem.
\end{center}

For a detailed account of those cases see\ \cite{gorr} or \cite{dok}. Some
of them were generalized through the addition of a gyrostatic moment \cite%
{gorr} and other potential and gyroscopic forces \cite{gorr}, \cite{yzamm-p}.

In the present article we aim at exploring the possibility of particular
solutions of the Bobylev-Steklov type for the problem of motion of a rigid
body about a fixed point in a field that generalizes the classical setting.
We assume that the body is acted upon by certain potential forces, which
admit symmetry axis fixed in space. The equations of motion for this problem
can be written in the Euler-Poisson form (e.g. \cite{lei}):

\begin{eqnarray}
A\dot{p}+(C-B)qr &=&\gamma _{2}\frac{\partial V}{\partial \gamma _{3}}%
-\gamma _{3}\frac{\partial V}{\partial \gamma _{2}},  \notag \\
B\dot{q}+(A-C)pr &=&\gamma _{3}\frac{\partial V}{\partial \gamma _{1}}%
-\gamma _{1}\frac{\partial V}{\partial \gamma _{3}},  \notag \\
C\dot{r}+(B-A)pq &=&\gamma _{1}\frac{\partial V}{\partial \gamma _{2}}%
-\gamma _{2}\frac{\partial V}{\partial \gamma _{1}},  \label{1}
\end{eqnarray}%
\begin{equation}
\dot{\gamma}_{1}+q\gamma _{3}-r\gamma _{2}=0,\dot{\gamma}_{2}+r\gamma
_{1}-p\gamma _{3}=0,\dot{\gamma}_{3}+p\gamma _{2}-q\gamma _{1}=0,  \label{2}
\end{equation}%
where $A,B,C$ are the principal moments of inertia, $p,q,r$ are the
components of the angular velocity of the body and $\gamma _{1},\gamma
_{2},\gamma _{3}$ are the components of the unit vector $\mathbf{\gamma }$
along the axis of symmetry of the force field, all being referred to the
principal axes of inertia at the fixed point. The potential $V$ depends only
on the Poisson variables $\gamma _{1},\gamma _{2},\gamma _{3}.$ In the
classical problem of a heavy body $V=a\gamma _{1}+b$ $\gamma _{2}+c\gamma
_{3}.$

Equations (\ref{1}) and (\ref{2}) admit three general first integrals: 
\begin{equation}
I_{1}=\frac{1}{2}Ap^{2}+\frac{1}{2}Bq^{2}+\frac{1}{2}Cr^{2}+V,\text{ the
energy integral}  \label{E}
\end{equation}

\begin{equation}
I_{2}=\gamma _{1}^{2}+\gamma _{2}^{2}+\gamma _{3}^{2}=1,\text{ the geometric
integral}  \label{g}
\end{equation}

\begin{equation}
I_{3}=Ap\gamma _{1}+Bq\gamma _{2}+Cr\gamma _{3},\text{ the areas integral}
\label{a}
\end{equation}

\section{A new solvable case}

The aim of this paper is to look for potentials $V$ which allow full
solution of the system (\ref{1}-\ref{2}) under the condition 
\begin{equation}
q=0  \label{3}
\end{equation}%
on the zero level of the areas integral, i.e.%
\begin{equation}
I_{3}=0  \label{4}
\end{equation}%
We assume that $A\neq C.$ without restriction on the third moment of inertia 
$B$. This choice of the problem is motivated by the classical
Bobylev-Steklov solution in the classical problem, which is characterized by
the same condition (\ref{3}) but with the potential $V=a\gamma _{1}$ and\
the additional restriction on the moments of inertia $A=2C.$

From (\ref{3}) and the middle equation of (\ref{1}) we get 
\begin{equation}
(A-C)pr-\gamma _{3}\frac{\partial V}{\partial \gamma _{1}}+\gamma _{1}\frac{%
\partial V}{\partial \gamma _{3}}=0  \label{5}
\end{equation}%
Differentiating this equality and using equations (\ref{1})-(\ref{4}) we
arrive at the expressions for the remaining angular velocities%
\begin{eqnarray}
p &=&-\sqrt{\frac{C\gamma _{3}}{A(A-C)\gamma _{1}}(\gamma _{1}\frac{\partial
V}{\partial \gamma _{3}}-\gamma _{3}\frac{\partial V}{\partial \gamma _{1}})}
\notag \\
r &=&\sqrt{\frac{A\gamma _{1}}{C(A-C)\gamma _{3}}(\gamma _{1}\frac{\partial V%
}{\partial \gamma _{3}}-\gamma _{3}\frac{\partial V}{\partial \gamma _{1}})}
\label{6}
\end{eqnarray}%
where $V$ satisfies the linear partial differential equation

\begin{eqnarray}
&&\gamma _{1}\gamma _{2}\gamma _{3}(A\frac{\partial ^{2}V}{\partial \gamma
_{1}^{2}}-C\frac{\partial ^{2}V}{\partial \gamma _{3}^{2}})+(A\gamma
_{1}^{2}+C\gamma _{3}^{2})(\gamma _{1}\frac{\partial ^{2}V}{\partial \gamma
_{2}\partial \gamma _{3}}-\gamma _{3}\frac{\partial ^{2}V}{\partial \gamma
_{1}\partial \gamma _{2}})  \notag \\
&&-\gamma _{2}(A\gamma _{1}^{2}-C\gamma _{3}^{2})\frac{\partial ^{2}V}{%
\partial \gamma _{1}\partial \gamma _{3}}  \notag \\
&&-(A-2C)\gamma _{2}\gamma _{3}\frac{\partial V}{\partial \gamma _{1}}%
+2(A-C)\gamma _{3}\gamma _{1}\frac{\partial V}{\partial \gamma _{2}}%
-(2A-C)\gamma _{1}\gamma _{2}\frac{\partial V}{\partial \gamma _{3}}=0
\label{e}
\end{eqnarray}

\bigskip On the other hand, from the first and third equations of (\ref{2}),
in virtue of (\ref{3}), we have%
\begin{equation}
\frac{d\gamma _{3}}{d\gamma _{1}}=\frac{\dot{\gamma}_{3}}{\dot{\gamma}_{1}}=%
\frac{C}{A}\frac{\gamma _{3}}{\gamma _{1}}  \label{7}
\end{equation}%
This can be readily integrated to give%
\begin{equation}
\gamma _{3}=\lambda \gamma _{1}^{C/A},\text{ \ \ \ }\lambda =const.
\label{8}
\end{equation}%
This means the trace of the vertical unit vector on the unit sphere attached
to the body lies on a cylindrical surface whose generators are parallel to
the $y-$ axis. Also, from (\ref{g}) one can express $\gamma _{2}$ as%
\begin{equation}
\gamma _{2}=\sqrt{1-\gamma _{1}^{2}-\lambda ^{2}\gamma _{1}^{\frac{2C}{A}}}
\label{9}
\end{equation}%
asd
Thus, five of the six Euler-Poisson variables are expressed in terms of the
last one: $\gamma _{1}.$ The relation with time can be determined by
separation of variables in the first Poisson equation. This finally gives%
\begin{equation}
t=\int \frac{d\gamma _{1}}{\sqrt{g(\gamma _{1})}}  \label{t}
\end{equation}%
where%
\begin{equation*}
g(\gamma _{1})=\frac{A\gamma _{1}^{\frac{A-C}{A}}}{C(A-C)\lambda }(1-\gamma
_{1}^{2}-\lambda ^{2}\gamma _{1}^{\frac{2C}{A}})(\gamma _{1}\frac{\partial V%
}{\partial \gamma _{3}}-\gamma _{3}\frac{\partial V}{\partial \gamma _{1}}%
)_{0}
\end{equation*}%
and $()_{0}$ in the right hand side means the value of the expression in
virtue of relations (\ref{8}, \ref{9}), so that $g$ is a function of the
single variable $\gamma _{1}$.

Summing up, we formulate the following

\begin{theorem}
For an arbitrary rigid body moving about a fixed point while acted upon by
forces with potential $V$ satisfying (\ref{e}), the Euler-Poisson equations (%
\ref{1}-\ref{2}) are solvable on the zero level of the areas integral under
the condition $q=0.$ The solution is parametrized in terms of $\gamma _{1}$
by expressions (\ref{6}, \ref{8}, \ref{9}) and relation with time is given
by (\ref{t}).
\end{theorem}

\section{The general form of the solution}

\bigskip It is not hard to construct the general solution of the linear PDE (%
\ref{e}), which may be written in the form%
\begin{equation}
V=V_{1}+V_{2},  \label{V}
\end{equation}%
where

\begin{eqnarray}
V_{1} &=&(A\gamma _{1}^{2}+C\gamma _{3}^{2})\int\limits^{A\gamma
_{1}^{2}+C\gamma _{3}^{2}}F(\frac{[A(\gamma _{1}^{2}+\gamma
_{3}^{2})-u]^{A/C}}{C(\gamma _{1}^{2}+\gamma _{3}^{2})-u})\frac{du}{u^{2}}
\label{V1} \\
V_{2} &=&(A\gamma _{1}^{2}+C\gamma _{3}^{2})G(\gamma _{2})  \label{V2}
\end{eqnarray}%
This form involves two arbitrary functions $F$ and $G,$ which we assume well
behaved on the poisson sphere (\ref{g}), so that all subsequent operations
on the potential $V$ are justified. With this form of the potential we can
rewrite the expressions for the Euler-Poisson variables in their final form
parametrized by $\gamma _{1}$ 
\begin{eqnarray}
(p,q,r) &=&(-\sqrt{\frac{2C}{A}}\lambda \gamma _{1}^{C/A},0,\sqrt{\frac{2A}{C%
}}\gamma _{1})\sqrt{\varpi }  \notag \\
(\gamma _{1},\gamma _{2},\gamma _{3}) &=&(\gamma _{1},\sqrt{1-\gamma
_{1}^{2}-\lambda ^{2}\gamma _{1}^{\frac{2C}{A}}},\text{ }\lambda \gamma
_{1}^{C/A})  \label{O}
\end{eqnarray}%
where%
\begin{eqnarray}
\varpi &=&\frac{F(-\lambda ^{\frac{2A}{C}}(A-C)^{\frac{A}{C}-1})}{A\gamma
_{1}^{2}+C\lambda ^{2}\gamma _{1}^{2C/A}}+\int^{A\gamma _{1}^{2}+C\lambda
^{2}\gamma _{1}^{2C/A}}F(\frac{[A(\gamma _{1}^{2}+\lambda ^{2}\gamma
_{1}^{2C/A})-u]^{A/C}}{C(\gamma _{1}^{2}+\lambda ^{2}\gamma _{1}^{2C/A})-u})%
\frac{du}{u^{2}}  \notag \\
&&+G(\sqrt{1-\gamma _{1}^{2}-\lambda ^{2}\gamma _{1}^{\frac{2C}{A}}}),
\end{eqnarray}%
and $\gamma _{1}$ is determined in terms of time by inverting the integral

\begin{equation}
t=\int \frac{d\gamma _{1}}{\sqrt{g(\gamma _{1})}},\text{ \ }g(\gamma _{1})=-2%
\frac{A\gamma _{1}^{2}}{C}(1-\gamma _{1}^{2}-\lambda ^{2}\gamma _{1}^{\frac{%
2C}{A}})\varpi  \label{T}
\end{equation}

It may now be verified that the solution given by the expressions (\ref{V}-%
\ref{T}) satisfies the Euler-Poisson equations (\ref{1}, \ref{2}). Also, one
can check that the total energy of the motion is in fact preserved and has
the value%
\begin{equation}
h=-F(-\lambda ^{\frac{2A}{C}}(A-C)^{\frac{A}{C}-1})
\end{equation}

\section{Generalization of the classical Bobylev-Steklov, Chaplygin and
Goriatchev cases}

It is difficult to foresee the explicit form of the part (\ref{V1}) of the
potential with a given choice of the function $F$ in the integrand. It may
also be impractical or even impossible to evaluate the integral in (\ref{V1}%
) for arbitrary $A$ and $C$ in closed form. In the classical Bobylev-Steklov
the moments of inertia are subject to a single condition $A=2C.$ We now
explore the form of the potential corresponding to certain simple forms of
the function $F$ under this same condition, to obtain a generalization of\
the classical Bobylev-Steklov case. It turns out that the sequences of
functions $\{u^{n+\frac{1}{2}}\},\{u^{-n}\}$ lead to rational potentials. As
an example we take the first three terms of each, so that%
\begin{equation}
F(u)=\sqrt{u}(A_{1}+A_{2}u+a_{3}u^{2})+\frac{B_{1}}{u}+\frac{B_{2}}{u^{2}}+%
\frac{B_{3}}{u^{3}}
\end{equation}%
gives the potential%
\begin{eqnarray}
V_{1} &=&a_{1}\gamma _{1}+a_{2}[8\gamma _{1}(\gamma _{1}^{2}+\gamma
_{3}^{2})+\frac{\gamma _{3}^{4}}{\gamma _{1}}]  \notag \\
&&+a_{3}[16\gamma _{1}(8\gamma _{1}^{4}+16\gamma _{1}^{2}\gamma
_{3}^{2}+9\gamma _{3}^{4})+\frac{16\gamma _{3}^{6}}{\gamma _{1}}+\frac{%
\gamma _{3}^{8}}{\gamma _{1}^{3}}]  \notag \\
&&+\frac{b_{1}}{\gamma _{3}^{2}}+b_{2}\frac{\gamma _{1}^{2}-\gamma _{3}^{2}}{%
\gamma _{3}^{6}}+b_{3}\frac{2\gamma _{1}^{4}-2\gamma _{1}^{2}\gamma
_{3}^{2}+\gamma _{3}^{4}}{\gamma _{3}^{10}}
\end{eqnarray}%
$a_{i},b_{i}$ being arbitrary parameters. If also we choose $G(\gamma
_{2})=C_{1}+C_{2}$ $\gamma _{2}^{2}$ and use the geometric relation (\ref{g}%
), we may write the full potential (\ref{V}) as%
\begin{equation}
V=V_{1}+c_{1}(\gamma _{1}^{2}-\gamma _{2}^{2})+c_{2}\gamma _{2}^{2}(2\gamma
_{1}^{2}+\gamma _{3}^{2})  \label{Vr}
\end{equation}

\bigskip The solution of the equations of motion corresponding to this
potential may be expressed by inserting it into (\ref{6}) and then
substituting expressions (\ref{8}, \ref{9}) for $\gamma _{3},\gamma _{2}.$
One can obtain $\gamma _{1}$ as a function of time by inverting the
hyperelliptic integral%
\begin{eqnarray}
t &=&\lambda ^{2}\sqrt{\frac{C}{2}}\int^{\gamma _{1}}\frac{dx}{\sqrt{g(x)}},
\\
g(x) &=&\frac{1}{x}(1-x^{2}-\lambda x)  \notag \\
&&\times \{\lambda ^{4}[2c_{2}x^{5}-2(4a_{2}-c_{2}\lambda
)x^{4}-2(2a_{2}\lambda +c_{1}+c_{2})x^{3}-(a_{1}-a_{2}\lambda ^{2})x^{2}] 
\notag \\
&&\qquad -2(b_{1}\lambda ^{2}+3b_{2})x+2b_{2}\lambda \}
\end{eqnarray}%
This integral becomes ultraelliptic when $c_{2}=0$ and elliptic if,
moreover, $a_{2}=b_{2}=0$. In the last case the whole solution may be
written as follows:

\begin{equation}
V=a_{1}\gamma _{1}+c_{1}(\gamma _{1}^{2}-\gamma _{2}^{2})+\frac{b_{1}}{%
\gamma _{3}^{2}}  \label{Vg}
\end{equation}%
\begin{eqnarray}
\gamma _{3} &=&\sqrt{\nu \gamma _{1}},\gamma _{2}=\sqrt{1-\nu x-x^{2}} 
\notag \\
p &=&-\sqrt{-\frac{1}{2C\nu \gamma _{1}}(2b_{1}+a_{1}\nu ^{2}\gamma
_{1}+2c_{1}\nu ^{2}\gamma _{1}^{2})},q=0,  \notag \\
r &=&\frac{1}{\nu }\sqrt{-\frac{2}{C}(2b_{1}+a_{1}\nu ^{2}\gamma
_{1}+2c_{1}\nu ^{2}\gamma _{1}^{2})}
\end{eqnarray}%
\begin{equation}
t=\sqrt{\frac{C}{2}}\nu \int^{\gamma _{1}}\frac{dx}{\sqrt{(x^{2}+\nu
x-1)(2b_{1}+\nu ^{2}a_{1}x+2\nu ^{2}c_{1}x^{2})}}  \label{tg}
\end{equation}%
Thus, the solution of the Euler-Poisson system (\ref{1}-\ref{2} ) with the
three-terms potential (\ref{Vg}) can be expressed explicitly,\ for the body
satisfying $A=2C$ and on the zero level of $I_{3}$, in terms of elliptic
functions of time$.$

It should be noted here that the the solution (\ref{Vg}, \ref{tg}) does not
depend on the second moment of inertia $B$, which remains arbitrary. If in
this solution we add the condition $B=A$, so that the the body has the
Kovalevskaya configuration $A=B=2C$, we obtain an intersection with three
well-known integrable problems of rigid body dynamics:

\begin{enumerate}
\item The full potential (\ref{Vg}) with three arbitrary parameters $%
a_{1},c_{1},b_{1}$ was pointed out first by Goriachev in \cite{gor}, but
only under the condition $A=B=2C$.

\item When $b_{1}=0,$ we get a problem of motion of a rigid body by inertia
in an ideal incompressible fluid (see e.g. \cite{ykov1})

\item $b_{1}=c_{1}=0,$Kovalevskaya's case of motion about a fixed point in
the uniform gravity field. This case is general integrable, i.e. on
arbitrary level of $I_{3}.$
\end{enumerate}

Formulas (\ref{tg}) give the explicit solutions of the first two cases under
the additional restriction $q=0$ and the third under the two restrictions $%
q=I_{3}=0,$ in terms of elliptic functions of time.

\section{\protect\bigskip Some closed-form and polynomial potentials}

Most problems of physical importance and all known completely integrable and
solvable cases in rigid body dynamics are characterized exactly or
approximately by potentials of simple forms that are mainly polynomial or
algebraic in the Poisson variables $\gamma _{1},\gamma _{2},\gamma _{3}$. It
is quite interesting to isolate possible solutions of polynomial or finite
form of the linear PDE (\ref{e}) satisfied by $V$ for arbitrary moments of
inertia. The part $V_{2}$ of (\ref{V}) is polynomial for any polynomial
choice of $G(\gamma _{2}).$ The part $V_{1}$ as expressed by (\ref{V1})
depends only on $\gamma _{1},\gamma _{3}$ and hence satisfies the following
particular version of equation (\ref{e}):%
\begin{eqnarray}
&&\gamma _{1}\gamma _{3}(A\frac{\partial ^{2}V}{\partial \gamma _{1}^{2}}-C%
\frac{\partial ^{2}V}{\partial \gamma _{3}^{2}})-(A\gamma _{1}^{2}-C\gamma
_{3}^{2})\frac{\partial ^{2}V}{\partial \gamma _{1}\partial \gamma _{3}} 
\notag \\
&&-(A-2C)\gamma _{3}\frac{\partial V}{\partial \gamma _{1}}-(2A-C)\gamma _{1}%
\frac{\partial V}{\partial \gamma _{3}}=0  \label{e1}
\end{eqnarray}%
We now assume a solution of this linear PDE as a combination of homogeneous
terms of the two variables $\gamma _{1},\gamma _{3}.$ After some trials to
normalize the singularities of the resulting differential equations we
arrive at the following form%
\begin{equation}
V=\sum\limits_{\nu }a_{\nu }\gamma _{1}^{\nu }g_{\nu }(v),\text{ \ }v=1+%
\frac{\gamma _{3}^{2}}{\gamma _{1}^{2}}  \label{v}
\end{equation}%
where the summation, with arbitrary constant coefficients $a_{\nu }$,
extends over the set of possible values of $\nu $ that will be determined
later.\emph{\ }This leads to the differential equations for each of the
functions $g_{\nu }$%
\begin{equation}
v(1-v)\frac{d^{2}g_{\nu }}{dv^{2}}+[(\nu -2+\frac{\nu }{2(\alpha -1)})v-%
\frac{\nu }{2}+2]\frac{dg_{\nu }}{dv}+(\frac{\alpha \nu }{\alpha -1}%
-2)g_{\nu }=0  \label{eg}
\end{equation}%
in which $\alpha =\frac{A}{C}.$ This is a hypergeometric equation whose
solutions for generic values of the parameters are%
\begin{eqnarray}
g_{\nu 1} &=&F(-\frac{\nu }{2},1-\frac{\alpha \nu }{2(\alpha -1)};-\frac{\nu 
}{2}+2;v),  \label{g1} \\
g_{\nu 2} &=&v^{\nu /2-1}F(-1,-\frac{\nu }{2(\alpha -1)};\frac{\nu }{2};v) 
\notag \\
&=&v^{\nu /2-1}\frac{\alpha -1+v}{\alpha -1}  \label{g2}
\end{eqnarray}%
The second solution $g_{\nu 2}$ contributes to (\ref{v}) a term%
\begin{equation}
A_{\nu }(\gamma _{1}^{2}+\gamma _{3}^{2})^{\nu /2-1}(\alpha \gamma
_{1}^{2}+\gamma _{3}^{2})=A_{\nu }(1-\gamma _{2}^{2})^{\nu /2-1}(\alpha
\gamma _{1}^{2}+\gamma _{3}^{2})
\end{equation}%
which is of the type (\ref{V2}), and can be better considered as included in
that form. It thus remains to consider the decomposition of the potential
part $V_{1}$ (\ref{V1})in the form implied by (\ref{V}) and (\ref{g1}). The
typical term will now read 
\begin{equation}
V_{1\nu }=\gamma _{1}^{\nu }F(-\frac{\nu }{2},1-\frac{\alpha \nu }{2(\alpha
-1)};-\frac{\nu }{2}+2;v)  \label{t1}
\end{equation}

\bigskip As we are looking for closed-form solutions we try to isolate cases
when the hypergeometric series terminates, giving a polynomial expression.
To that end, we note that the hypergeometric function satisfies the relation%
\begin{equation*}
F(a,b;c;v)=(1-v)^{c-a-b}F(c-a,c-b,c)
\end{equation*}%
and thus 
\begin{equation*}
F(-\frac{\nu }{2},1-\frac{\alpha \nu }{2(\alpha -1)};-\frac{\nu }{2}%
+2;v)=(1-v)^{1+\frac{\alpha \nu }{2(\alpha -1)}}F(2,1+\frac{\nu }{2(\alpha
-1)};-\frac{\nu }{2}+2;v)
\end{equation*}%
There are 3 obvious possible types of such closed-form expressions, in which
the hypergeometric series truncates:

1) $a$ is a negative integer, while $c$ is not a negative integer greater
than $-n$. This gives only one non-trivial possibility $\nu =2$, which leads
to potential of the type (\ref{V2}) for a constant $G(\gamma _{2}).$

2) $b$ is a negative integer $-n$, while $c$ is not a negative integer
greater than $-n$. This gives the sequence $\{\nu =\frac{2(\alpha -1)}{%
\alpha }(n+1),$ $n=0,1,2,...\}.$ The third index for this choice becomes $%
\alpha +1+(\alpha -1)n$, which is positive for all positive $n.$ This choice
invokes an infinite sequence of terms in the potential. The typical term in
this case has the form%
\begin{equation}
\frac{\gamma _{1}^{2n}}{\gamma _{3}^{2(\alpha n+\alpha -1)}}F(2,-n\text{ }%
;\alpha +1+(\alpha -1)n;1+\frac{\gamma _{3}^{2}}{\gamma _{1}^{2}})
\label{t2}
\end{equation}%
The numerator of this term is a homogeneous polynomial of degree $2n$ in the
two variables $\gamma _{1}$ and $\gamma _{3}$. The power $2(\alpha n+\alpha
-1)$ in the denomenator is always positive. This sequence of potential terms
is undefined at $\gamma _{3}=0$ for all $n.$

3) $c-b$ is a negative integer $-n,$ while $\alpha $ is not one of the $n+1$
rational numbers $\frac{n+1}{n-1-k},$ $k=0,1,...,n.$ This leads to the
sequence $\{\nu =-2(\alpha -1)(n+1),$ $n=0,1,2,...\}$ and to potentials of
the type%
\begin{equation}
\gamma _{1}^{\frac{2(\alpha -1-n)}{\alpha }}\cdot \gamma _{1}^{2n}F(-n,-%
\frac{\alpha -1}{\alpha }(n+1);1-n+\frac{n+1}{\alpha };1+\frac{\gamma
_{3}^{2}}{\gamma _{1}^{2}})  \label{t3}
\end{equation}%
The product of the second and third factors in this expression is a
polynomial of degree $2n.$ The behaviour of the first term differs depends
on the two numbers $\alpha $ and $n.$

The final form of the potential becomes

\bigskip 
\begin{eqnarray*}
V &=&(A\gamma _{1}^{2}+C\gamma _{3}^{2})\sum_{n}A_{n}\gamma _{2}^{n} \\
&&+\gamma _{1}^{\frac{2(\alpha -1)}{\alpha }}\sum_{n}B_{n}\gamma
_{1}^{2n(1-1/\alpha )}F(-n,-\frac{\alpha -1}{\alpha }(n+1);1-n+\frac{n+1}{%
\alpha };1+\frac{\gamma _{3}^{2}}{\gamma _{1}^{2}}) \\
&&+\frac{1}{\gamma _{3}^{2(\alpha -1)}}\sum_{n}\frac{C_{n}}{\gamma
_{3}^{2\alpha n}}\gamma _{1}^{2n}F(2,-n\text{ };\alpha +1+(\alpha -1)n;1+%
\frac{\gamma _{3}^{2}}{\gamma _{1}^{2}})
\end{eqnarray*}

\end{document}